\documentclass[a4paper,11pt]{article}
\pdfoutput=1 

\usepackage{jheppub} 

\usepackage[T1]{fontenc} 

\title{On a Modified Klein-Gordon Equation with Vacuum-Energy Contributions}

\author[a,1]{Dor Gabay,\note{Corresponding author.}}
\author[b]{Sijo K. Joseph}

\affiliation[a]{Department of Physical Electronics, Tel-Aviv University, Tel-Aviv 69978, Israel.}
\affiliation[b]{Quantum Gravity Research, Topanga, CA 90290, USA.}

\emailAdd{dorgabay@post.tau.ac.il,sijo@quantumgravityresearch.org}

\abstract{We define a modified covariant Klein-Gordon (KG) equation containing quantum vacuum contributions arising from the self-interaction of matter with its own internal kinetic energy. 
The modified KG equation is exemplified for a variety of vacuum fields and various properties of the equation are articulated thereof. 
Generalized commutation and Energy-Momentum relations are characterized for a null vacuum-phase scenario of the proposed vacuum field $\lambda$. 
Within this limited scenario, a representation theorem is introduced suggesting that one can equally modify the spacetime structure or momentum operator in articulating the proposed quantum theory. 
Such a modified KG equation is further shown to eliminate infrared and the ultraviolet divergences in the generalized Klein-Gordon propagator.}

\begin{document} 
\maketitle
\flushbottom

\section{Introduction}
Geometrical interpretations of quantum mechanics have attracted much attention in recent years~\cite{QMGeometry,our_cosmo_paper}.
Many theories are attempting to unravel a spacetime structure which could conform with the underlying laws of quantum mechanics~\cite{Sidharth_QM,QRGM_Susskind}. 
Given our understanding of gravity as a pure classical field theory on a curved manifold, the notion of a geometrical theory of quantum mechanics deserves special consideration. 
In quantum theory, matter is defined on a flat spacetime structure and the equations of motion (i.e. Klein-Gordon) disregard any interaction the matter may have with its own internal kinetic energy. 

For this reason, Bohmian theory~\cite{BohmI,BohmII} deserves special attention; its geometrical interpretation makes it a unique candidate for coupling both theories geometrically.
In such a framework, one can associate the density with particle trajectories by separating the wavefunction into a real-valued density and phase~\cite{PHollandBook,SheldonReview}.
Recently it has been identified that the relativistic Bohmian quantum potential $Q$ can play a significant role in defining quantum matter fields within the framework of conformal gravity ~\cite{Shojai_Article,ShojaiBohmianQM}. 
Shojai et al.~\cite{Shojai_Article,ShojaiCnstrAlgebra,ShojaiBohmianQM,Shojai_ScalarTensor} proposed that the exponential form of the conformal factor must be assumed within the Bohmian framework to avoid tachyonic behavior. Our studies have indicated that these kinds of extensions of the conformal factor make corrections to quantum theory in the very fundamental level. 
The unexplored exponential form of the conformal factor brings about interesting notions of physics which have yet to be explored.
It is therefore always fascinating to study these corrections in the quantum framework and discern their implication on equations of motion, commutation relations, and renormalization procedures in quantum field theory. 

Furthermore, in considering a conformal theory of gravity, Manheim and Bender have suggested that the Hermitian nature of quantum mechanics can be generalized to PT-Symmetric Hamiltonians~\cite{Bender1,Bender2,Bender3,Bender4,Manheim_PTSym}.
Spacetime reflection have been shown to generate real-eigenvalues for a broader class of Hamiltonians, giving a rich physical description of nature beyond Hermiticity. 
By considering a conformal theory of gravity, with a conformal factor characterized by the particle's internal kinetic energy, one can more easily incorporate accelerative contributions within a quantum mechanical framework. 

In our previous work ~\cite{our_cosmo_paper}, we used a conformal factor composed of the quantum potential, along with a constraint, to bypass Weinberg's no-go theorem~\cite{WeinbegCosmo89} and introduce a scalar vacuum field expression using the following action 

\begin{eqnarray}
A[g_{\mu\nu},{\Omega}, S, \rho, \lambda]&=&
\frac{1}{2\kappa}\int{d^4x\sqrt{-g}\left(R\Omega^2-6\nabla_{\mu}\Omega\nabla^{\mu}\Omega\right)}  \nonumber \\
& & +\int{d^4x\sqrt{-g} \left(\frac{\rho}{m}\Omega^2 \nabla_{\mu}S \nabla^{\mu}S-m\rho\Omega^4\right)} \nonumber \\
& & +\int{d^4x\sqrt{-g}\lambda\left[\ln{\Omega^2}-\left(\frac{\hbar^2}{m^2}\frac{\nabla_{\mu}\nabla^{\mu}\sqrt{\rho}}{\sqrt{\rho}}
\right)\right]} \label{Actioneq}. 
\end{eqnarray}
The alleged self-interaction, resulting from the conformal transformation, evidently allowed us to properly define a classical description of matter with gravity for a significantly small mass $m$. 
A Lagrange multiplier $\lambda$ was used to constraint the conformal factor to the exponent of the quantum potential $e^Q$. 
In equating the scalar curvature equation to the trace of Einstein's equation, the resulting scalar field equation, composed of the Lagrange multiplier, was used to effectively bypass Weinberg's no-go theorem. 
The Lagrange multiplier $\lambda$ was further suggested to be characteristic of the vacuum-energy due to its dominating nature for considerably small masses and its embedding of Heisenberg's uncertainty principle (e.g. to linear-order of the conformal factor)\cite{our_cosmo_paper}. 
It is by no means suggested that the proposed geometrical theory would describe gravity itself, but it is our hope that it would bring us one step closer to fundamentally understanding the underlying nature of quantum mechanics, in particular, how accelerative contributions might play a role in such an extended theory, whereby the interaction of a particle with its own internal kinetic energy is considered. 

In this manuscript, we address the importance of considering the interaction of a particle with its own kinetic energy. 
The conformal factor alleviates the point-like nature of the particle, and the interaction of matter's kinetic energy with spacetime becomes unavoidable. A modified second-order Klein-Gordon (KG) equation, containing both the KG field $\psi$ and a newly introduced vacuum field $\lambda$~\cite{our_cosmo_paper}, is mapped from the Bohmian to wavefunction picture. 
An interesting dissipative contribution, composed of the quantum force and current density, arises as a consequence of a newly introduced gauge connection, and speculations of its physical meaning are made thereof. 
At first, the modified KG equation is explored whence the density and phase of the vacuum $\lambda$ are aligned with that of the KG field $\lambda=\rho e^{iS/\hbar}=\sqrt{\rho}\psi$.
It is further shown that, when the phase associated to the vacuum field vanishes (e.g. null-vacuum phase scenario) and the density \textit{alone} aligns with the density of the KG field $\lambda=\rho$, the acclaimed vacuum field contributions can more fundamentally be described as \textit{non-local} quantum corrections within the energy-momentum and commutation relations. 
A representation theorem follows, whereby a flat-space or curved-space interpretation of the modified Klein-Gordon equation can be articulated either from an extended momentum operator or differential form operator, respectively. 
Finally, implications are made to renormalization procedures, where it is shown that infrared and ultraviolet divergences are alleviated for the extended commutation relations.
    
\section{Modified Klein-Gordon Equation}
We re-express the Bohmian equation of motion arising from the conformally transformed action (Eq.~\ref{Actioneq}) into its more pleasant wavefunction form. 
In the remainder of this section, we assume the exponential constraint within the action to be of linear-order (i.e. conformal factor $\Omega^2\approx 1+Q$). 
Diving straight into the exponential form of the conformal factor $\Omega^2$ is not trivial and will instead be explored in future works. 
Whether equations of motion are studied within the Bohmian or wavefunction framework should not significantly matter. 
Typically, the wavefunction form is preferred due its inclusion of the continuity equation and the equation of 
motion into a single field equation.
From our previous work~\cite{our_cosmo_paper}, the Bohmian equation of motion and continuity equation, without vacuum field contributions, can be expressed in terms of the density $\sqrt{\rho}$, classical action $S$, and conformal factor $\Omega^2$
\begin{eqnarray}
\nabla_{\mu}S\nabla^{\mu}S - m^2\Omega^2=0 \label{EoMnoVac}\\
\nabla_{\mu}(\rho\Omega^2\nabla^{\mu}S)=0 \label{EoMcnteqn}
\end{eqnarray}
Here, $\Omega^2=1+\frac{\hbar^2}{m^2} \frac{\nabla_{\mu} \nabla^{\mu}\sqrt{\rho}}{\sqrt{\rho}}$ is the conformal factor to linear-order. 
With some effort, one can map this equation to its wavefunction form by setting $\sqrt{\rho}=\sqrt{\psi^*\psi}$ and 
\begin{eqnarray}
\nabla^{\mu}S = -i\hbar\Bigl(\frac{\nabla^{\mu}\psi}{\psi}-\frac{\nabla^{\mu}\sqrt{\rho}}{\sqrt{\rho}}\Bigr) \\
\nabla^{\mu}S = \frac{-i\hbar}{2}\Bigl(\frac{\nabla^{\mu}\psi}{\psi}-\frac{\nabla^{\mu}\psi^*}{\psi^*}\Bigr) \label{Sexprsn}
\end{eqnarray}
Even with these relations, retrieving the wavefunction equation from Eq.~\ref{EoMnoVac}-~\ref{EoMcnteqn} is not a trivial task.
After some tedious formulation, the continuity equation can be re-organized to a form which can be substituted into the equation of motion
\begin{eqnarray}
\nabla_{\mu}S\nabla^{\mu}S = -\hbar^2\Biggl(\frac{\nabla_{\mu}\nabla^{\mu}\psi}{\psi} 
- \frac{\nabla_{\mu}\nabla^{\mu}\sqrt{\rho}}{\sqrt{\rho}} 
-\frac{\nabla_{\mu}\Omega^2}{\Omega^2}\Bigl(\frac{\nabla^{\mu}\sqrt{\rho}}{\sqrt{\rho}} 
-\frac{\nabla^{\mu}\psi}{\psi}\Bigr)\Biggr) \nonumber \label{EqmDaSDas}.
\end{eqnarray}
By substituting our results into Eq.~\ref{EoMnoVac} and expanding the density in terms of the wavefunction and its conjugate, a generalized Klein-Gordon equation is obtained
\begin{eqnarray}
\Box\psi+\frac{1}{2}\frac{\nabla_{\mu}\Omega^2}{\Omega^2}\Bigl(\frac{\nabla^{\mu}\psi}{\psi}-\frac{\nabla^{\mu}\psi^{*}}{\psi^{*}}\Bigr)\psi+ \frac{m^2}{\hbar^2}\psi=0 \label{WFeqnp1} 
\end{eqnarray}
One could further substitute the current density 
$J^{\mu}=\frac{-i\hbar}{2}\Bigl(\psi^{*}\nabla^{\mu}\psi-\psi\nabla^{\mu}\psi^{*}\Bigr)=\rho\nabla^{\mu}S$ into the equation of motion to obtain a compact representation
\begin{eqnarray}
\Box\psi+\frac{i}{\hbar}\Bigl(\frac{\nabla_{\mu}\Omega^2\,J^{\mu}}{\Omega^2\rho}\Bigr)\psi+ \frac{m^2}{\hbar^2}\psi=0. 
\label{WFeqn1} 
\end{eqnarray}
The component $\frac{\nabla_{\mu}\Omega^2}{\Omega^2}$ can be perceived as the quantum force for an exponential constraint 
($\Omega^2=e^{Q}$) of the conformal factor $\frac{\nabla_{\mu}\Omega^2}{\Omega^2}= \nabla_{\mu}\ln{\Omega^2}=\nabla_{\mu}Q$. 
Therefore, $\frac{\nabla_{\mu}\Omega^2}{\Omega^2}$ in Eq.~\ref{WFeqn1} is  a linear-order approximation of $\nabla_{\mu}Q$, and in this case suffices.
The coupling between the self-generated quantum force $\nabla_{\mu}Q$ and current density $J^{\mu}$ 
breaks current conservation (i.e $\nabla_{\mu}J^{\mu}\neq 0$). 
The current density $J^{\mu}$ appears to provide a feedback to itself, obeying the relation $\nabla_{\mu}J^{\mu}=-\nabla_{\mu}Q\,J^{\mu}$ derived from the continuity equation (Eq.~\ref{EoMcnteqn}). 
It is therefore apparent that the modified continuity equation (See Eq.~\ref{EoMcnteqn}) can essentially be substituted into the conformally transformed Klein-Gordon equation, allowing one to express Eq.~\ref{WFeqn1} more generally as 
\begin{eqnarray}
\Box\psi-\frac{i}{\hbar}\Bigl(\frac{\nabla_{\mu}J^{\mu}}{\rho}\Bigr)\psi+ \frac{m^2}{\hbar^2}\psi=0. \label{QWFeqn1} 
\end{eqnarray}
Here, the wavefunction equation contains an accelerative contribution resulting from the implied quantum force  
$i\,\hbar(\frac{\nabla_{\mu}\Omega^2\,J^{\mu}}{\Omega^2\rho})$. 
This term can best be described as the particle conforming to its background kinetic energy, via a feedback interaction, by dissipating energy associated to $J_{\mu}\nabla^{\mu}Q$. A relevant question is: if energy is being dissipated in Eq.~\ref{QWFeqn1} by the quantum force, where is it emerging from in the first place? Keep this question in mind as we progress forward. 

To better understand the meaning of Eq.~\ref{QWFeqn1}, we elaborate further on the Bohmian interpretation of the wavefunction $\psi=\sqrt{\rho}e^{iS/\hbar}$. Firstly, a relationship can be determined for $\nabla_{\mu}\sqrt{\rho}$
\begin{eqnarray}
\nabla_{\mu}\sqrt{\rho}=\sqrt{\rho}\Bigl(\frac{\nabla_{\mu}\psi}{\psi}-\frac{i}{\hbar}\nabla_{\mu}S\Bigr) \label{duRho}.
\end{eqnarray}
Here, $S$ is simply the phase associated to the scalar field $\psi$. One can further take the second derivative of $\sqrt{\rho}$ to obtain
\begin{eqnarray}
\frac{\nabla_{\mu}\nabla^{\mu}\sqrt{\rho}}{\sqrt{\rho}}&=&\nabla_{\mu}\Bigl(\frac{\nabla^{\mu}\psi}{\psi}-\frac{i}{\hbar}
\nabla^{\mu}S\Bigr)+\Bigl(\frac{\nabla^{\mu}\psi}{\psi}-\frac{i}{\hbar}\nabla^{\mu}S\Bigr)^2. \nonumber \\ 
\label{duduRho}
\end{eqnarray}
After some tedious algebra, one finds an interesting relationship for the quantum potential $Q$ and the scalar field $\psi$ 
\begin{eqnarray}
Q\psi=\Bigl(\nabla_{\mu}-\frac{i}{\hbar}\frac{J_{\mu}}{\rho}\Bigr)\Bigl(\nabla^{\mu}-\frac{i}{\hbar}\frac{J^{\mu}}{\rho}\Bigr)\psi=\mathcal{D}_{\mu}\mathcal{D}^{\mu}\psi. \label{Qpsi}
\end{eqnarray}
Here, $\mathcal{D}_{\mu}=(\nabla_{\mu}-\frac{i}{\hbar}\frac{J_{\mu}}{\rho})$, where $\rho$ and $J_{\mu}$ are representative of the charge and 4-vector current densities, respectively.
An operator representation of the quantum potential $Q$ on the wavefunction $\psi$ can therefore be defined as $\hat{Q}=\mathcal{D}_{\mu}\mathcal{D}^{\mu}$.

Furthermore, by expanding the gauge connection of Eq.~\ref{Qpsi}, one can re-express Eq.~\ref{QWFeqn1} in terms of the covariant derivative $\mathcal{D}_{\mu}=(\nabla_{\mu}-\frac{i}{\hbar}\nabla_{\mu}S)=(\nabla_{\mu}-\frac{i}{\hbar}\frac{J_{\mu}}{\rho})$ 
\begin{eqnarray}
\mathcal{D}_{\mu}\mathcal{D}^{\mu}\psi=0, \label{QWFeqnFlatDalph} 
\end{eqnarray}
Proving that the proposed gauge connection is inherently connected to the dissipative term found in Eq.~\ref{WFeqn1}. 
When the current density $J_{\mu}$ is orthogonal to the quantum force $\nabla_{\mu}Q$, such a gauge connection is removed, and the original KG equation is retrieved. 
Furthermore, the mass $m$ is conspicuously hidden within the classical term $\nabla_{\mu}S\nabla^{\mu}S$ of the contracted covariant derivatives $\mathcal{D}_{\mu}\mathcal{D}^{\mu}$, conforming with our original assumption of the energy-momentum relation of a classical particle~\cite{our_cosmo_paper}. In considering this fact, one can reach an obvious, yet peculiar, relation
\begin{eqnarray}
\frac{1}{\hbar^2}\nabla_{\mu}S\nabla^{\mu}S=\frac{1}{\hbar^2}\frac{J_{\mu}J^{\mu}}{\rho^2}=\frac{m^2}{\hbar^2} \to \frac{J_{\mu}J^{\mu}}{\rho^2}=m^2. \label{current_cple}
\end{eqnarray}
Here, the mass seems to directly correspond to the coupled form of the current-density. Under the immature assumption that mass is a manifestation of the self-interaction of the current density, one can speculate that the particle may be interacting with a yet unexplored field variable. Going back to the previously posed question (why should the particle's energy arbitrarily dissipate?), one plausible answer is that another field variable should accompany Eq.~\ref{WFeqn1} once accelerative contributions are accounted for. Allowing the particle to arbitrarily dissipate energy with no manifestation of an underlying source would violate energy conservation. It is therefore highly compelling to define a field which could constraint the particle's kinetic energy and implicitly act as a source for such dissipative behavior. 

What if the vacuum-energy simply arises as a consequence of balancing matter's expanding and contracting forces (i.e. accelerative contributions) of spacetime? 
In enforcing such a balancing mechanism, we were intrigued at the notion of $\lambda$ being characteristic of the vacuum-energy; especially after identifying consistent trends of the cosmological constant with the modified Einstein's equation~\cite{our_cosmo_paper}. 
By introducing the acclaimed vacuum field $\lambda$, the aforementioned problem of arbitrarily dissipating energy could be alleviated. 
More specifically, that dissipative forces within Eq.~\ref{QWFeqnFlatDalph} could be balanced out by quantum vacuum contributions. 
The complete form of the modified Bohmian equation of motion containing the alleged vacuum field (presented in~\cite{our_cosmo_paper}), can be defined along with the continuity equation in Eq.~\ref{EoMcnteqn}
\begin{equation}
\nabla_{\mu}S \nabla^{\mu}S-m^2\Omega^2+\frac{\hbar^2}{2m 
	\Omega^2\sqrt{\rho}}\Bigl[\Box{\Bigl(\frac{\lambda}{\sqrt{\rho}}\Bigr)}-\lambda\frac{\Box\sqrt{\rho}}{\rho}\Bigr]=0. \label{EqMotion} 
\end{equation}
Here, the procedure for defining the wavefunction equation is not so different from before. After substituting Eq.~\ref{EqmDaSDas} into Eq.~\ref{EqMotion} and using the continuity equation (Eq.~\ref{EoMcnteqn}) to simplify $\frac{\nabla_{\mu}\Omega^2\,J^{\mu}}{\Omega^2}=-\nabla_{\mu}J^{\mu}$, we obtain a wavefunction equation coupled to the quantum vacuum 
\begin{equation}
\Bigl(\Box+\frac{m^2}{\hbar^2}\Bigr)\psi-\frac{i}{\hbar}\Bigl(\frac{\nabla_{\mu}J^{\mu}}{\rho}\Bigr)\psi 
- \frac{1}{2m\Omega^2\sqrt{\rho}}\Bigl[\Box\Bigl({\frac{\lambda}{\sqrt{\rho}}}\Bigr)
- \lambda\frac{\Box{\sqrt{\rho}}}{\sqrt\rho}\Bigr]\psi =0 \label{WFeqnVacuum}. 
\end{equation}
The vacuum density $\lambda$ (in the linear-order case $\Omega^2=1+Q$) obeys a first order differential equation and is a nontrivial function of the density $\sqrt{\rho}$  
\begin{eqnarray}
\frac{m^2}{\hbar^2}\lambda=\nabla_{\mu}\Bigl(\lambda\frac{\nabla^{\mu}\sqrt{\rho}}{\sqrt{\rho}} \Bigr) \label{LambdaNiceEq}
\end{eqnarray}
Here, $m^2/\hbar^2$ is simply the Compton wavelength. Equation ~\ref{WFeqnVacuum} contains second-order covariant derivatives of the density and wavefunction simultaneously. Using the gauge connection proposed in Eq.~\ref{Qpsi} and substituting Eq.~\ref{LambdaNiceEq} into Eq.~\ref{WFeqnVacuum}, one can further simplify the equation of motion
\begin{eqnarray}
\Omega^2\mathcal{D}_{\mu}\mathcal{D}^{\mu}\psi 
- \Bigl[\frac{1}{2m\rho}\Bigl(\Box
- \frac{2m^2}{\hbar^2}\Bigr)\lambda\Bigr]\psi =0 \label{WFeqnVacuumF}. 
\end{eqnarray}
From Eq.~\ref{WFeqnVacuumF} and Eq.~\ref{LambdaNiceEq}, it can be seen that $\lambda$ acts as an effective feedback mechanism, deeming it possible to deal with seemingly simple coupled second-order differential equations (Eq.~\ref{WFeqnVacuum}).

Articulating two fields simultaneously can prove to be difficult in interpreting the underlying physics. There are various special scenarios which can be imposed on the added field $\lambda$ to give the coupled equations a much richer interpretation. 
The simplest of these is by enforcing that the variation of $\lambda$ with respect to $\nabla_{\mu}\sqrt{\rho}$ in Eq.~\ref{LambdaNiceEq} vanishes ($\delta\lambda/\delta\nabla_{\mu}\sqrt{\rho}=0$), followed by applying a partial derivative $\nabla_{\mu}$ on both sides
\begin{eqnarray}
\Box\lambda=2\lambda\nabla_{\mu}\Bigl(\frac{\nabla^{\mu}\sqrt{\rho}}{\sqrt{\rho}}\Bigr)+2\nabla_{\mu}\lambda\frac{\nabla^{\mu}\sqrt{\rho}}{\sqrt{\rho}} \label{lambdaEq0}
\end{eqnarray}
Given the zeroth and second order variations of the action with respect to the density $\sqrt{\rho}$ were already taken, such a procedure is reminiscent of removing all variations of the vacuum field with respect to the density 
\begin{eqnarray}
\frac{\delta\lambda}{\delta\sqrt{\rho}}=\frac{\delta\lambda}{\delta\nabla_{\mu}\sqrt{\rho}}=\frac{\delta\lambda}{\delta\Box\sqrt{\rho}}=0 \label{Const0}
\end{eqnarray}
Expanding Eq.~\ref{LambdaNiceEq} and substituting it into Eq.~\ref{lambdaEq0} gives a density-independent $\lambda$ equation
\begin{eqnarray}
\Bigl(\Box-\frac{2m^2}{\hbar^2}\Bigr)\lambda=0 \label{LambdaEqF}
\end{eqnarray}
Here, $\lambda$ appears to take a tachyonic form. In such a constrained class of solutions, henceforth called the `null vacuum-phase scenario,' Eq.~\ref{WFeqnVacuumF} further simplifies to the usual KG equation
\begin{eqnarray}
\Bigl(\Box+\frac{m^2}{\hbar^2}\Bigr)\psi=0 \label{WFeqnVacuumF0}. 
\end{eqnarray}
Here, the vacuum contribution is removed and the dissipative contribution along with it. 
The removal of $\nabla_{\mu}Q J^{\mu}$ can be justified by recognizing that the imposed condition in Eq.~\ref{lambdaEq0} nulls the quantum potential $\nabla_{\mu}Q=0$. This can be shown by mapping $\lambda\to\Omega^2\lambda$ and, in a similar fashion, taking the variation of the $\lambda$-equation with respect to $\delta/\delta\nabla_{\mu}\sqrt{\rho}$. Undergoing such a procedure is equivalent to changing the constraint within the action of Eq.~\ref{Actioneq} from $\lambda\left[\ln{\Omega^2}-\left(\frac{\hbar^2}{m^2}\frac{\nabla_{\mu}\nabla^{\mu}\sqrt{\rho}}{\sqrt{\rho}}
\right)\right] \to \lambda\left[\Omega^2-e^{\frac{\hbar^2}{m^2}\frac{\nabla_{\mu}\nabla^{\mu}\sqrt{\rho}}{\sqrt{\rho}}
}\right]$.

Eq.\ref{WFeqnVacuumF0} no longer contains any inherent feedback contribution associated to the vacuum field $\lambda$. 
The fact that $\lambda$ no longer appears in the wavefunction equation, does not mean it does not effect it in some way. 
The same result can be reached by setting $\lambda=\lambda_0\rho$, where $\lambda_0$ is some yet unknown unitless constant associated to the vacuum field. 
One can verify this by simply recognizing that Eq.~\ref{LambdaEqF} can be reproduced by setting $\lambda=\rho$.
This can be conceptually depicted as the quantum mechanical density ($\rho=|\psi|^2$) conforming with the vacuum. 
Intuitively, this suggests the usual KG equation only holds under the assumption of: 1) a null vacuum-phase; and 2) that, regardless of the KG equation, the density associated to the wavefunction is manifestly tachyonic (Eq.~\ref{LambdaEqF}). The static solutions associated to Eq.~\ref{LambdaEqF} in flat-space can be articulated as (for thoroughness, we assume $c\neq1$ here)
\begin{eqnarray}
\hbar^2c^2\nabla^2\lambda+\Bigl(2m^2c^4+E^2\Bigr)\lambda=0 \label{LambdaEqFr}
\end{eqnarray}
Here, $E$ is the energy associated to the vacuum. For convenience, we assign a variable for the bracketed term in Eq.~\ref{LambdaEqFr}; $\beta=(2m^2c^4+E^2)/(\hbar^2 c^2)$. There seem to be three possible solutions to Eq.~\ref{LambdaEqFr}: (1) 
$\nabla^2\lambda=0$ for $\beta=0$; 
(2) $\nabla^2\lambda+|\beta|\lambda=0$ for $\beta>0$; (3) $\nabla^2\lambda-|\beta|\lambda=0$ for $\beta<0$. 
$\lambda$ in 
(1) obeys Laplace's equation and satisfies hyperbolic sine and cosine functions, 
(2) conforms to the solution of the Helmholtz equation, and (3) procures the solution of the homogeneous screened Poisson equation.
We further assume that any energy $E$ greater than the Compton frequency would violate energy conservation. This leaves (1) and (2) to be the regimes of most practical interest. It is important to remember that Eq.~\ref{LambdaEqFr} applies to a specific subclass of solutions imposed by the null vacuum-phase scenario $\lambda=\rho$. 
By only studying a subset of solutions to Eq.~\ref{LambdaNiceEq}, we can get a better understanding of the tachyonic nature of the added field $\lambda$, and more importantly, identify its strong resemblance to the Helmholtz equation for $\beta>0$ (Eq.~\ref{LambdaEqFr}). 

In its more generalized form, one can remove the assumption of a null vacuum-phase and instead assume that the vacuum phase conforms with that of the KG field $\lambda=\rho e^{iS/\hbar}=\sqrt{\rho}\psi$. In such a scenario, the two-field equation can be written purely in terms of KG field $\psi$ and the vacuum field coupling no longer vanishes as in Eq.~\ref{WFeqnVacuumF0}. By substituting the phase-dependent $\lambda$ into Eq.~\ref{WFeqnVacuum}, one finds that the KG equation can be expressed as
\begin{eqnarray}
\mathcal{D}_{\mu}\mathcal{D}^{\mu}\psi+ \frac{e^{iS/\hbar}}{2m\Omega^2}\Bigl(\mathcal{D}_{\mu}\mathcal{D}^{\mu}-\Box\Bigr)\psi =0.\label{EoMphase}
\end{eqnarray}
Here, the vacuum contributions are analogous to the phase components of the gauge connection in Eq.~\ref{Qpsi}
\begin{eqnarray}
\Bigl(\mathcal{D}_{\mu}\mathcal{D}^{\mu}-\Box\Bigr) = \Bigl(\frac{i}{\hbar}\frac{\nabla_{\mu}Q J^{\mu}}{\rho}+\frac{1}{\hbar^2}\frac{J_{\mu}J^{\mu}}{\rho^2}\Bigr) = \Bigl(-\frac{i}{\hbar}\frac{\nabla_{\mu}J^{\mu}}{\rho}+\frac{m^2}{\hbar^2}\Bigr) \label{vac_phase}
\end{eqnarray}
Where we assume $\frac{1}{\hbar^2}\frac{J_{\mu}J^{\mu}}{\rho^2}$ is proportional to $\frac{m^2}{\hbar^2}$, as suggested by Eq.~\ref{current_cple}. 
From Eq.~\ref{vac_phase}, it appears that the $\lambda$-contribution in Eq.~\ref{EoMphase} contributes a feedback to the KG field $\psi$ proportional to $e^{iS/\hbar}/2m\Omega^2$. Unlike the scale symmetry breaking implicated by the mass term within the usual KG equation, the breaking here (Eq.~\ref{EoMphase}) seems to be a consequence of the vacuum contribution 
\begin{eqnarray}
\Box\psi-\frac{i}{\hbar}\frac{\nabla_{\mu}J^{\mu}}{\rho}\psi+\frac{m^2}{\hbar^2}\psi+ \frac{e^{iS/\hbar}}{2m\Omega^2}\Bigl(-\frac{i}{\hbar}\frac{\nabla_{\mu}J^{\mu}}{\rho}+\frac{m^2}{\hbar^2}\Bigr)\psi =0.\label{EoMphase_v0a}
\end{eqnarray}
The factor $e^{iS/\hbar}/2m\Omega^2$ governs how `strongly' the symmetry is broken.
In a hypothetical scenario whereby $e^{iS/\hbar}/2m\Omega^2\approx -1$, Eq.~\ref{EoMphase} reduces to $\Box\psi=0$, making it scale-invariant in a flat-space scenario. 
Unlike the scale-symmetry breaking $m^2$ of the usual KG-equation, $e^{iS/\hbar}/2m\Omega^2$ embarks upon a much more intuitive descriptor, one which enforces the scale-symmetry breaking to be inversely proportion to the mass rather than directly proportional to the mass squared. 
Therefore, as the mass $m\to\infty$, the proposed symmetry breaking mechanism slowly diminishes $\frac{e^{iS/\hbar}}{2m\Omega^2}(\mathcal{D}_{\mu}\mathcal{D}^{\mu}-\Box)\to0$, and, along with it, the effect of the vacuum energy to the particle dynamics. In the large-mass regime (where $\psi$ begins to be highly localized), the mass contribution $m^2\psi$ begins to dominate, the dissipation term associated to the quantum force becomes negligible (i.e. $\nabla_{\mu}Q\to0$), and the propagation term $\hbar^2\Box\psi$ becomes insignificant.

The phase component $e^{iS/\hbar}$ is particularly interesting. It articulates some form of phase-dependence of the particle to the vacuum field. To better understand it, we explore an alternative fashion of Eq.~\ref{EoMphase} when $\lambda=\rho e^{-iS/\hbar}=\sqrt{\rho}\psi^{*}$
\begin{eqnarray}
\Omega^2\mathcal{D}_{\mu}\mathcal{D}^{\mu}\psi+ \frac{e^{iS/\hbar}}{2m}\Bigl(\mathcal{\widetilde{D}}_{\mu}\mathcal{\widetilde{D}}^{\mu}-\Box\Bigr)\psi^{*} =0 \label{EoMphase_v1}. 
\end{eqnarray}
Here, $\mathcal{\widetilde{D}_{\mu}}=(\nabla_{\mu}+\frac{i}{\hbar}\nabla_{\mu}S)$ is simply the conjugate of $\mathcal{D}_{\mu}=(\nabla_{\mu}-\frac{i}{\hbar}\nabla_{\mu}S)$. The negative phase $e^{-iS/\hbar}$ of $\psi^{*}$ seems to balance out the positive phase of $\psi$ in the last component, leaving no inherent phase within the vacuum contribution as in Eq~\ref{EoMphase}. Although, this result is deceiving. The resemblance between Eq.~\ref{EoMphase} and Eq.~\ref{EoMphase_v1} is more obvious once one takes the following relation into account
\begin{eqnarray}
\psi^{*}\mathcal{D}_{\mu}\mathcal{D}^{\mu}\psi=\psi^{*}(Q\psi)=(Q\psi^{*})\psi=(\mathcal{\widetilde{D}}_{\mu}\mathcal{\widetilde{D}}^{\mu}\psi^{*})\psi \label{QPsiRel_v0}
\end{eqnarray}
With this relation accounted for, the two equations of motion can be articulated as conjugates of one another, suggesting that there is a strict independence in the directionality of the phase between the fields $\lambda$ and $\psi$. This implies that any property associated to Eq.~\ref{EoMphase} still persists, regardless of the directionality of the phase of $\lambda$. 

Reverting back to the assumption of $\lambda=\rho e^{iS/\hbar}$ (for the remainder of this paper), the corresponding $\lambda$-equation is modified from its former expression (Eq.~\ref{LambdaNiceEq}). In its density form, it can be written as 
\begin{eqnarray}
\Box\rho+\frac{i}{\hbar}\nabla_{\mu}S\nabla^{\mu}\rho-\frac{2m^2}{\hbar^2}\rho=0 \label{rhoEq} \label{lambdaPhase}
\end{eqnarray}
Here, an additional phase contribution enters the vacuum expression. In the null vacuum-phase $J_{\mu}\nabla^{\mu}\rho=0$, suggesting that, in such a scenario, an orthogonality exists between the density and energy-momentum $\nabla_{\mu}S$ analogous to the orthogonality of the aforementioned quantum force and energy-momentum $\nabla_{\mu}Q\nabla^{\mu}S=0$. It is important to remind the reader that, in deriving Eq.~\ref{LambdaEqF}, the vacuum phase was assumed to be null. Eq.~\ref{lambdaPhase} is therefore a generalized version of Eq.~\ref{LambdaEqF} (vacuum phase $e^{iS/\hbar}$ does not have to be null for Eq.~\ref{lambdaPhase} to equate to Eq.~\ref{LambdaEqF}).

One can also express the $\lambda$-equation purely in terms of the KG field $\psi$, with the necessary gauge connection $\mathcal{D}_{\mu}$, by substituting $\lambda=\rho e^{iS/\hbar}=\sqrt{\rho}\psi$. After some algebraic manipulation
\begin{eqnarray}
\psi^{*}\mathcal{D}_{\mu}\mathcal{D}^{\mu}\psi+\mathcal{\widetilde{D}}_{\mu}\psi^{*}\nabla^{\mu}\psi=\frac{m^2}{\hbar^2}\psi^{*}\psi. \label{lambdaPsi_v0}
\end{eqnarray}
The difficulty in interpreting Eq.~\ref{lambdaPsi_v0} is in its simultaneous dependence on the field $\psi$ and $\psi^{*}$. The field $\psi$ is inseparable from its conjugate pair and, furthermore, an asymmetry appears in the gauge connection of the two (left side, last component). Eq.~\ref{lambdaPsi_v0} can be further simplified to its null vacuum-phase form (e.g. $\lambda=\rho$) by adding the conjugate of Eq.~\ref{lambdaPsi_v0} to itself
\begin{eqnarray}
\psi^{*}\mathcal{D}_{\mu}\mathcal{D}^{\mu}\psi+\nabla_{\mu}\psi^{*}\nabla^{\mu}\psi=\frac{2m^2}{\hbar^2}\psi^{*}\psi. \label{lambdaPsi_v1}
\end{eqnarray}
One can alternatively substitute $\lambda=\psi^{*}\psi$ in Eq.~\ref{LambdaEqF} to get an alternative, but equivalent form of Eq.~\ref{lambdaPsi_v1}
\begin{eqnarray}
\psi^{*}\mathcal{D}_{\mu}\mathcal{D}^{\mu}\psi+\mathcal{\widetilde{D}}_{\mu}\psi^{*}\mathcal{D}^{\mu}\psi=\frac{m^2}{\hbar^2}\psi^{*}\psi. \label{lambdaPsi_v1a}
\end{eqnarray}
Here, it can be shown that $\mathcal{\widetilde{D}}_{\mu}\psi^{*}\mathcal{D}^{\mu}\psi=\nabla_{\mu}\psi^{*}\nabla^{\mu}\psi-\frac{m^2}{\hbar^2}\psi^{*}\psi$. In such a null vacuum-phase scenario, whereby Eq.~\ref{EoMphase} reduces to Eq.~\ref{WFeqnVacuumF0}, one can further substitute the relation $\psi^{*}\mathcal{D}_{\mu}\mathcal{D}^{\mu}\psi=0$ into Eq.~\ref{lambdaPsi_v1a} (or Eq.~\ref{lambdaPsi_v1})
\begin{eqnarray}
\mathcal{\widetilde{D}}_{\mu}\psi^{*}\mathcal{D}^{\mu}\psi=\frac{m^2}{\hbar^2}\psi^{*}\psi. \label{lambdaPsi_v2}
\end{eqnarray}
Unlike the gauge connection in the phase-dependent $\lambda$-equation ($\widetilde{D}_{\mu}\psi^{*}\nabla^{\mu}\psi$), the $\psi$-form of the null vacuum-phase scenario is symmetric in the gauge connection $\widetilde{D}_{\mu}\psi^{*}\mathcal{D}^{\mu}\psi$. The transition from an asymmetric to symmetric form of the $\lambda$-equation is only made possible when the vacuum-phase disappears. A question of key importance is whether the dynamics governing the current density of the wavefunction equation in the null vacuum-phase scenario have drastically changed. The answer is No. Even though the equations governing the wavefunction have essentially changed, current density conservation is still satisfied. One can exemplify this by substituting Eq.~\ref{WFeqnVacuumF0} into Eq.~\ref{lambdaPsi_v2} 
\begin{eqnarray}
\nabla_{\mu}J^{\mu}=\frac{i\hbar}{2}(\psi\Box\psi^{*}-\psi^{*}\Box\psi)=0 \label{CE_v0}
\end{eqnarray}
Therefore, it is only in the limit of $\lambda=\rho$, that one returns to the usual Klein-Gordon field with $\nabla_{\mu}J^{\mu}=0$. Beyond this limit, the gauge connection $\mathcal{D}_{\mu}$ can shine some light at the role of the particle's changed reference frame (i.e. current density implies a reference frame). In applying the gauge connection to the probability current density, an effective 'velocity' of the particle is obtained
\begin{eqnarray}
J_{\mu}^{\mathcal{D}} =-\frac{i\hbar}{2}(\psi^{*}\mathcal{D}_{\mu}\psi-\psi\mathcal{\widetilde{D}}_{\mu}\psi^{*}) =J_{\mu}-\rho\nabla_{\mu}S=0 \label{CE_v1}
\end{eqnarray}
Here, the current density $J_{\mu}=\rho\nabla_{\mu}S$. By the result of Eq.~\ref{CE_v1}, the current density $J_{\mu}^{\mathcal{D}}$ is always zero. This intuitively makes sense: the gauge connection effectively removes the phase contribution from $\nabla_{\mu}\psi$; further suggesting that the proposed gauge connection is analogous to placing the particle within its comoving frame! The physical essence of the comoving frame can be better understood by considering the implication of Weinberg's no-go theorem~\cite{WeinbegCosmo89}. In the null vacuum-phase scenario $\mathcal{D}_{\mu}\mathcal{D}^{\mu}=0$ the particle essentially embeds itself in an accelerative frame of the vacuum field. Beyond this scenario $\mathcal{D}_{\mu}\mathcal{D}^{\mu}\neq0$, the particle is not exactly within the described null accelerated frame, rather in a reference frame implicated by the vacuum-energy $\mathcal{D}_{\mu}\mathcal{D}^{\mu}=f(\lambda)$, where $f(\lambda)$ is the newly considered vacuum contribution to the particle dynamics.

\section{Revised Commutation Relations}
There is an interesting implication to the null vacuum-phase scenario, whereby the vacuum aligns itself with the density of the KG field. In the limit of $\lambda\to\rho$ (whereby the phase does not \textit{totally} diminish), the two fields, $\lambda$ and $\psi$, can be written into a single field expressed as an infinite-order PDE. To achieve this, we transition our analysis to the exponential constraint of Eq.~\ref{Actioneq} (e.g. whereby $\Omega^2=e^{Q}$). The familiar KG field should now contain a nonlocal dependence 
\begin{eqnarray}
\mathcal{D}_{\mu}\mathcal{D}^{\mu}\psi=0 \label{WFeqnVacuumF1}. 
\end{eqnarray}
\begin{eqnarray}
\mathcal{D}_{\mu}\mathcal{D}^{\mu}\psi+\frac{1}{2}\frac{\nabla_{\mu}\psi^{*}}{\psi^{*}}\nabla^{\mu}\psi-\frac{m^2}{\hbar^2}\psi=0 \label{lambdaPsi_v3}
\end{eqnarray}
Here, in transitioning from a linear to exponential constraint, an additional $\mathcal{D}_{\mu}\mathcal{D}^{\mu}\psi$ is added to the left side of Eq.~\ref{lambdaPsi_v3}. 
We remind the reader that Eq.~\ref{LambdaNiceEq}, Eq.~\ref{WFeqnVacuumF}, Eq.~\ref{LambdaEqF} and Eq.~\ref{rhoEq} were studied for the linear-order constraint ( $\Omega^2=1+Q$ ). These equations can be generalized to the exponential constraint simply by replacing $m^2\to(1-Q)m^2$, in the corresponding $m^2 \lambda$ terms. 
After careful analysis, it is apparent that Eq.~\ref{lambdaPsi_v3} takes the form of a diffusion equation acting on the KG field. As a consequence, the translation operator $e^{-\gamma^{\mu}\nabla_{\mu}}$ contains the following property for a uniform vector field $\gamma_{\mu}$
\begin{eqnarray}
e^{-\gamma^{\mu}\nabla_{\mu}}\psi(x^{\mu})=\psi(x^{\mu}-\gamma^{\mu}), \label{shiftTheorem}
\end{eqnarray}
Here, $\gamma^{\mu}$ acts as the shift vector of the internal variable $x_{\mu}$. In understanding this shift property, the extra field equation for $\psi$ can be re-expressed as a non-local operator acting on Eq.~\ref{WFeqnVacuumF1}. In the case where $\gamma_{\mu}$ is no longer uniform, the shift operation of Eq.~\ref{shiftTheorem} no longer holds. For arbitrary $\gamma_{\mu}$, we can state that the middle term in Eq.~\ref{lambdaPsi_v3} acts as a more general operator $\hat{O}$ acting on $x_{\mu}$
\begin{eqnarray}
e^{-\gamma^{\mu}(x)\nabla_{\mu}}\psi(x^{\mu})=\psi(\hat{O}x^{\mu}), \label{GenOperTheorem}
\end{eqnarray}
When $\gamma^{\mu}(x)$ is a constant vector, the internal variable operator $\hat{O}$ acts as a simple translation in the spacetime variable ($\hat{O}x^{\mu}\to x^{\mu}-\gamma^{\mu}$). 
When $\gamma^{\mu}(x)=\alpha x^{\mu}$, the operator $\hat{O}$ acts instead as a scaling operator on the spacetime ($\hat{O}x^{\mu}\to e^{\alpha}x^{\mu}$).

As a result, using the expression for the general operator given in Eq.\ref{GenOperTheorem}, the aforementioned wavefunction (Eq.~\ref{WFeqnVacuumF1}) and vacuum (Eq.~\ref{lambdaPsi_v3}) equations can be combined into a single equation of motion. 
The diffusion equation given in Eq.~\ref{lambdaPsi_v3} has already been analyzed by several authors in the context of p-adic string theory~\cite{Padic_MathPhys,Calcagni2010,RollingTachyon}. 
In the referenced theories, a diffusion equation effectively localizes a nonlocal exponential operator by imposing a shift mechanism on the internal variable of the scalar field.

Taking these previous works into consideration, Eq.~\ref{lambdaPsi_v3} can be rewritten as
\begin{eqnarray}
\left(\frac{2h^2}{m^2}\mathcal{D}_{\mu}\mathcal{D}^{\mu} -2 \right)\psi=-\gamma^{\mu}(x)\nabla_{\mu}\psi, \label{lambdaPsi_v4}
\end{eqnarray}
where $\gamma^{\mu}(x)=\frac{h^2}{m^2}\frac{\nabla_{\mu}\psi^{*}}{\psi^{*}}$. The corresponding transformation on the internal spacetime variable can then be imposed on the KG field $\psi$
\begin{eqnarray}
e^{\left(\frac{2h^2}{m^2}\mathcal{D}_{\mu}\mathcal{D}^{\mu} -2 \right)}\psi(x_{\mu})=e^{-\gamma^{\mu}\nabla_{\mu}}\psi(x^{\mu}) =\psi(\hat{O}x^{\mu})\label{lambdaPsi_v5}
\end{eqnarray}
Considering the fact the exponential operator only acts on the internal spacetime variable $\psi(x^{\mu})\to \psi(\hat{O}x^{\mu})$, Eq.~\ref{WFeqnVacuumF1} can be re-written as
\begin{eqnarray}
\mathcal{D}_{\mu}\mathcal{D}^{\mu}\Bigl(e^{(\frac{2h^2}{m^2}\mathcal{D}_{\mu}\mathcal{D}^{\mu}-2)}\psi(x^{\mu})\Bigr)=0 \label{WFeqnVacuumF2}. 
\end{eqnarray}
Hence,
\begin{eqnarray}
e^{\frac{2h^2}{m^2}\mathcal{D}_{\mu}\mathcal{D}^{\mu}}\mathcal{D}_{\mu}\mathcal{D}^{\mu}\psi=0 \label{WFeqnVacuumF3}. 
\end{eqnarray}
It is to be noted that Eq.~\ref{WFeqnVacuumF3} is an infinite order partial differential equation which needs an extra localization condition to make complete physical sense; Eq.~\ref{lambdaPsi_v3} provides just such a condition. 
In the usual quantum theory, this 'diffusive' behavior is missing. Only in considering the back-reaction of a particle to its own internal kinetic energy, via a scalar-tensor theory, were we able to retrieve an expression for the diffusive equation articulated in Eq.~\ref{lambdaPsi_v3}. 
In this section our sole purpose is to show how the quantum theory alone gets modified in the operator sense, independent of the diffusion equation.

It can easily be seen that Eq.~\ref{WFeqnVacuumF3} heavily modifies the quantum mechanical commutation relations.
Although Heisenberg's uncertainty principle has long been the underlying assumption of a very successful theory~\cite{QMSourceBook}, it can be argued that higher-order contributions are needed to satisfy a generalized form of the commutation relations. 
To exemplify this, let us first start by handling the linear-order extension of the theory. 
The canonical commutation relations corresponding to the uncertainty principle for a tensor metric $g_{\mu\nu}$ with the metric signature $(+,-,-,-)$ can more generally be defined as
\begin{eqnarray}
[\hat{x}_{\mu},\hat{p}_{\nu}]=i\hbar g_{\mu\nu}.
\end{eqnarray}
To first-order, the conformal factor $\Omega^2=1+\frac{\hbar^2}{m^2}\frac{\nabla_\alpha \nabla^\alpha\sqrt{\rho}}{\sqrt{\rho}}=1+Q$, can be used to express the more generalized commutation relations on a flat-space background ${g}_{\mu\nu}=\Omega^2 \eta_{\mu\nu}$. 
In assuming the existence of a yet unknown momentum operator $\hat{P}_\nu$, the modified commutation relation can then be expressed as
\begin{eqnarray}
[\hat{x}_{\mu},\hat{P}_{\nu}]=i\hbar g_{\mu\nu}=i\hbar \Omega^2 \eta_{\mu\nu}=i\hbar (1+Q) \eta_{\mu\nu}.. \label{comRel0}
\end{eqnarray}
Given the above equation, an appropriate question is whether such a functional form of the conformal factor can be replaced by an operator? The answer is Yes. Using the gauge connection found in Eq.~\ref{Qpsi}, Eq.~\ref{comRel0} can be re-expressed into an operator form 
\begin{eqnarray}
[\hat{x}_{\mu},\hat{P}_{\nu}]\psi=i\hbar \eta_{\mu\nu} \Bigl(1+\frac{\hbar^2}{m^2}\mathcal{D}_\alpha \mathcal{D}^\alpha \Bigr)\psi.
\end{eqnarray}
Here, the gauge connection $\mathcal{D}_\alpha \mathcal{D}^\alpha\psi$ replaces $Q\psi$. The generalized commutation relation can then be written in terms of a gauge-connection induced momentum operator $\hat{p}_\alpha=i\hbar \mathcal{D}_{\alpha}$
\begin{eqnarray}
[\hat{x}_{\mu},\hat{P}_{\nu}]=i\hbar \eta_{\mu\nu} \Bigl(1-\frac{1}{m^2} \hat{p}_\alpha \hat{p}^\alpha \Bigr),
\end{eqnarray}
And the \textit{overall} modified momentum operator $\hat{P}_\mu$ can be further articulated as a function of $\hat{p}_\mu$
\begin{eqnarray}
\hat{P}_\mu = \Bigl(1-\frac{1}{m^2} \hat{p}_\alpha \hat{p}^\alpha \Bigr)\hat{p}_\mu \label{CRfirst}
\end{eqnarray}
Eq.~\ref{CRfirst} assumes a quantum potential to first-order of the conformal factor. 
Higher order contributions of the momentum operator can be obtained by considering the single, infinite-order scalar field equation corresponding to Eq.~\ref{WFeqnVacuumF3}. 
Since $\hat{Q}={\mathcal{D}_\alpha \mathcal{D}^\alpha}$ commutes with itself, it can be shown that the exponential function 
$\Omega^2=e^{Q}$ in the differential geometric theory directly corresponds to an exponential operator 
$e^{\frac{\hbar^2}{m^2} \mathcal{D}_\alpha \mathcal{D}^\alpha}$ within the null vacuum-phase scenario
\begin{eqnarray}
[\hat{x}_{\mu},\hat{P}_{\nu}]=i\hbar\Omega^2 \eta_{\mu\nu}=i\hbar \eta_{\mu\nu} e^{\frac{\hbar^2}{m^2}\mathcal{D}_\alpha \mathcal{D}^\alpha} ~\label{ComRel0}
\end{eqnarray}
\begin{eqnarray}
[\hat{x}_{\mu},\hat{P}_{\nu}]=i\hbar \eta_{\mu\nu} (1+\frac{\hbar^2}{m^2}\mathcal{D}_\alpha \mathcal{D}^\alpha)(1+\frac{\hbar^2}{m^2}\frac{\mathcal{D}_\alpha \mathcal{D}^\alpha}{2})... 
\end{eqnarray}
Here, the exponential operator has been broken into fragments characterizing consecutively smaller length scales of spacetime. The non-tachyonic form of the conformal factor $\Omega^2=e^{-\frac{1}{m^2} \hat{p}_\alpha \hat{p}^\alpha}$ can similarly be expanded into a series
\begin{eqnarray}
\hat{P}_\mu =\Bigl(1 - \frac{1}{m^2} \hat{p}_\alpha \hat{p}^\alpha + \frac{1}{2!m^4} (\hat{p}_\alpha \hat{p}^\alpha)^2 - ... 
\Bigr)\hat{p}_\mu. \label{GenPmu}
\end{eqnarray}
Here, the proposed momentum operator conforms with the aforementioned infinite-order scalar field equation (Eq.~\ref{WFeqnVacuumF3}). The modified, infinite-order momentum operator can be seen as a physical consequence of matter's self-interaction with its own internal kinetic energy. As noted earlier, such infinite-order corrections are considered in p-adic string theory~\cite{Padic_MathPhys,Calcagni2010,RollingTachyon}.

The corrections appearing in the momentum operator $\hat{P}_{\mu}$ are proportional to $\frac{\hbar^2}{m^2}$, deeming lower-order contributions more important than their higher-order counterparts.
The smaller the mass $m$, the more significant the higher-order contributions.
One can then ask: why can such higher-order corrections only be considered within the framework of momentum operators?
The answer is: they can be considered in either framework (probablistic/geometric) depending on one's preference. 
To better grasp this, we further explore differential forms and their corresponding operators within a geometrical framework. In the argument that follows, it is the length scale, rather than the momentum operator, which matters in articulating the higher-order contributions. 
By analyzing the infinitesimal length element $ds^2$ and fixing the momentum operator to its ordinary flat-space form, one can 
see that the spacetime naturally allows for deformations which become more significant on smaller scales
\begin{eqnarray}
ds^2=dX_{\mu}dX^{\mu}=\Omega^2 \eta_{\mu\nu}dx^{\mu}dx^{\nu}
\end{eqnarray}
\begin{eqnarray}
dX_{\mu}dX^{\mu}\sqrt{\rho}=dx_{\mu}dx^{\mu}\Bigl(1+\frac{\hbar^2}{m^2}\nabla_{\alpha}\nabla^{\alpha}\Bigr)\sqrt{\rho}
\end{eqnarray}
Here, ${dX}^{\mu}$ is the differential form. By further taking the square root of both sides and presuming the approximation 
$\Bigl(1+\frac{\hbar^2}{m^2}\nabla_{\alpha}\nabla^{\alpha}\Bigr)^{\frac{1}{2}}\approx 
\Bigl(1+\frac{\hbar^2}{2m^2}\nabla_{\alpha}\nabla^{\alpha}\Bigr)$,
\begin{eqnarray}
d\hat{X}^{\mu}=dx^{\mu}\Bigl(1+\frac{\hbar^2}{2m^2}\nabla_{\alpha}\nabla^{\alpha}\Bigl), \label{difFormOpr0}
\end{eqnarray} 
we arrive at the linear approximation of what we call the \textit{differential form operator} $d\hat{X}^{\mu}$. 
The exponential conformal factor $\Omega^2=e^{Q}$ can be considered in a similar fashion to Eq.~\ref{GenPmu} with a key difference: the differential form operator $d\hat{X}^{\mu}$ must be confined to second order to satisfy bilinear form of the metric, and therefore must act on the density $d\hat{X}^{\mu}\sqrt{\rho}=dX^{\mu}\sqrt{\rho}$,
\begin{eqnarray}
dX_{\mu}dX^{\mu}= \eta_{\mu\nu}{dx}^{\mu}{dx}^{\nu}(1+\frac{\hbar^2}{m^2}Q)(1+\frac{\hbar^2}{m^2}\frac{Q}{2})...  \label{difform_v0}
\end{eqnarray}
Here, the continuity equation, inherently contained within the guage connection $\mathcal{D}_{\mu}\mathcal{D}^{\mu}$, is now contained within the quantum potential $Q$, via Eq.~\ref{Qpsi}. 
The corresponding momentum $p^\mu$ now takes on the classical form $p^\mu=m\frac{d\hat{X}^{\mu}}{d\tau}$, where $\tau$ is simply the affine parameter associated to the manifold. 
In the spatial representation, $d\hat{X}^{\mu}$ and $p^\mu$ are both real quantities. They have to be treated as two separate entities, $d\hat{X}^{\mu}$ defines the differential manifold while $p^\mu$ articulates the geodesic equation.

It is important to note that we are not claiming that the curvature associated to the differential form of Eq.~\ref{difform_v0} is gravity itself! It is simply the feedback contribution of the particle to its own quantum mechanical kinetic energy. 
Our hope is that, in considering a purely quantum mechanical entity, accelerative contributions considered within General Relativity can be further discerned within a generalized theory of quantum-gravity. 
Whether gravity is the result of such a coupling needs to be further explored. 
Regardless, one can define a representation theorem for the incorporation of such accelerative contributions by modifying the quantum mechanical commutation relations in one of two ways:
\begin{eqnarray}
(\hat{x}_\mu,\hat{P}_\nu)\iff(d\hat{X}_\mu,p_\nu).
\end{eqnarray}
Such a representation theorem suggests that one can either consider the modified momentum operator in a flat-spacetime, or adopt the differential form operator in characterizing a curved-spacetime structure. 
Here, a `Flat-spacetime' does not imply that gravity cannot be incorporated as an affine-connection in the covariant derivatives of the modified KG equation, it is simply meant to indicate the manner in which the uncertainty, implicated by the commutation relation, is embedded within a manifold structure.
By adopting the differential form operator $d\hat{X}_{\mu}$, one is lead to deal with the differential geometric framework, where no complex numbers are encountered. 
By otherwise considering a modified momentum operator $\hat{P}_{\mu}$, the flat spacetime may naturally lead to a more intuitive  representation of the corresponding equations of motion. 
Both are formally equivalent in the null vacuum-phase scenario. In the section which follows, we consider the energy-momentum relations and propagators within the flat-space representation $(\hat{x}_\mu,\hat{P}_\nu)$, to elucidate the modified nature of the wavefunction equation.

\section{Associated Energy-Momentum Relations}
A modern interpretation of Energy-Momentum relations can be defined by the system's energy, rest mass, and momentum $E^2=m^2+p^2$. 
The potential unification of U(1) symmetric quantum and accelerative fields (i.e. gravitational fields) raises a conceptually luring question: Does the classical relation, by which we base our theories, need to be modified to align with unified theories? 

By accounting for the generalized momentum operator $P_{\mu}$ (See Eq.~\ref{GenPmu}) within the energy-momentum relation $P_{\mu}P^{\mu}+m^2=0$, different orders of the commutation relations corresponding to Eq.~\ref{ComRel0} can be represented
\begin{eqnarray}
E^2=p^2+m^2  + \frac{2}{m^2} (p_{\mu}p^{\mu})^2 - \frac{2^2}{2!m^4} (p_{\mu}p^{\mu})^3 + \frac{2^3}{3!m^6} (p_{\mu}p^{\mu})^4 + ... \label{EPeq}
\end{eqnarray}
By taking only the first-order correction $(p_{\alpha}p^{\alpha})^2$ of Eq.~\ref{EPeq}, 
the modified energy-momentum relation now contains higher-order spatial and temporal contributions
\begin{equation}
E^2=p^2+m^2-\frac{4}{m^2}{E}^{2}{p}^{2}+\frac{2}{m^2}{p}^{4}+\frac{2}{m^2}{{E}^{4}} \label{EPeq0}.\\
\end{equation}
The additional components can be interpreted as a feedback mechanism (i.e. self-interaction) resulting from the transformation operator $\hat{O}$ of Eq.~\ref{lambdaPsi_v5}. 
These self-interactions, to first-order, have been shown to account for vacuum energy corrections (i.e. Lamb shift~\cite{SidharthLambRevw}) in quantum system.
In the nonrelativistic framework, the proposed relations are given by
\begin{equation}
E=m + \frac{p^2}{2m} + \frac{1}{m^3} (p^2-E^2)^2 
- \frac{2}{2!m^5} (p^2-E^2)^3 + \frac{2^2}{3!m^7} (p^2-E^2)^4 - ...
\end{equation}
Here, the higher-order energy-momentum contributions are effectively dampened by $1/2m$, but can nonetheless play a prior role in characterizing the effect of the vacuum on particle dynamics. 

We can further exemplify the consequence of the relation in Eq.~\ref{EPeq0} to particle-dynamics by studying propagators within renormalization procedures. 
Using the exponential form of the momentum operator in Eq.~\ref{GenPmu}, the modified Klein-Gordon equation can be expressed in a similar manner to the infinite-order scalar field equation found in Eq.~\ref{WFeqnVacuumF3} 
\begin{eqnarray}
\hat{P}_\mu\hat{P}^\mu \psi =0 \implies \hbar^2\,e^{\frac{2\hbar^2}{m^2}\mathcal{D}_{\mu}\mathcal{D}^{\mu}}\,\mathcal{D}_{\mu}\mathcal{D}^{\mu}\psi=0. \label{InfOrderEoM}
\end{eqnarray}
Using the basic definition of the propagator $\tilde{G}(k)$
\begin{eqnarray}
\Bigl(\hbar^2\,e^{\frac{2\hbar^2}{m^2}\mathcal{D}_{\mu}\mathcal{D}^{\mu}}\,\mathcal{D}_{\mu}\mathcal{D}^{\mu}\Bigr) G(x,x')=-i\delta^{4}(x-x'),
\end{eqnarray}
its spectral representation can be further articulated 
\begin{eqnarray}
\tilde{G}(k)=\frac{i\,\exp{\Bigr(-\frac{2}{m^2}(-\hbar^2 k^2+m^2-i\hbar\tilde{\epsilon}(k))\Bigl)}}{(- m^2 + i\hbar\tilde{\epsilon}(k)+\hbar^2 k^2)}.
\end{eqnarray}
Here,  ${\epsilon}(k)=\frac{1}{(2\pi)^4}\int{\Bigl(\frac{\nabla_{\mu}J^{\mu}}{\rho}\Bigr)e^{ik_{\mu}x^{\mu}}d^4x}$ and $\tilde{\epsilon}(k)=[{\epsilon}(k)*\tilde{G}(k)]\tilde{G}^{-1}(k)$. $\tilde{\epsilon}$ is an imaginary contribution resulting from the gauge connection in Eq.~\ref{Qpsi}. $\epsilon(k)$ must convolve with $\tilde{G}(k)$ to properly characterize $\tilde{\epsilon}(k)$. The imaginary contribution $\tilde{\epsilon}$ within the Green's function seems to contain $\tilde{G}$, suggesting a looping mechanism is required to fully characterize the Green's function. We emphasize that such a behavior is a consequence of the self-interaction of matter with its own internal kinetic energy. $\tilde{\epsilon}$ is unique in that it effectively shifts the poles away from the real axis as is done in fixing the Feynman propagator. 
It can be easily seen that the propagator eliminates the UV- and IR-Divergences associated to Feynman diagrams.
The exponential in the denominator will always have higher-order powers of $k$ to counter 
balance the power terms appearing in the numerator for any loop order. 
Similarly, the imaginary component will preserve a finite value in the denominator for $k\to 0$. 
Although the higher-order contributions appearing in the aforementioned commutation and energy-momentum relations lead to seemingly interesting generalizations of the equation of motion, higher-orders can prove to be disadvantageous both conceptually and numerically. 
The main criticism concerned with the higher-order derivative theories is the Ostrogradsky instabilities associated to the equations. 
In string field theories, non-local equations with infinitely many powers of the d'Alembertian operator are frequently studied~\cite{Freund87padic,Zwiebach_String2002,Gianluca2008}. 
One of the main issues with higher-order derivate theories is that the system Hamiltonian linearly depends on some of the momentum coordinates, allowing the momentum to freely take on negative values. Negative values of momentum lead to Hamiltonians unbounded from below, resulting in the infamous Ostrogradsky instability~\cite{OstroInstab2015}. 
Ostrogradsky instability theorem states that ``\textit{For any non-degenerate theory whose dynamical variable is higher than second-order in time derivative there exists a linear instability}''~\cite{Woodard2007Ostro,Woodard2015Ostro}. 
This creates issues like negative norm states or ghost states in the corresponding quantum theory. 
Mannheim and Bender proposed that PT-Symmetric Hamiltonians~\cite{Bender1,Bender2,Bender3,Bender4,Manheim_PTSym,BenderManheimGhost08} can effectively remove such ghosts from higher-order derivative theories. 
Others have also shown that one can exorcise the ghosts, and hence eliminate such instabilities, by using constraints to reduce the dimensionality of the phase-space~\cite{Chen2013ostro,BarnabyKamranOstro}. 

Still, a question arises here: can the alleged instability associated to higher-order derivative theories be directly applied to infinite-order equations? The answer is ``No.'' Recently, N. Barnaby and N. Kamran have indicated that equations with infinitely many derivatives can never be consistently viewed as the $N\to \infty$ limit of some $N$-th order equation~\cite{BarnabyKamranOstro}. They have also shown that differential equations of infinite-order do not generically admit infinitely many initial data. There is a crucial difference between finite-order and infinite-order derivative theories; the former acts locally on the field variable while the latter nonlocally on the field variable. Therefore, one should be careful in rejecting infinite-order theories simply on the basis of Ostrogradsky instability. 

At this point, we should remind the reader that the obtained commutation and energy momentum relations apply only to the null vacuum-phase scenario, whereby the vacuum and quantum matter formally disassociate. Further work needs to be done to understand the physical essence of matter-vacuum interactions beyond the null vacuum-phase scenario. One elegant example has been demonstrated whereby the vacuum field lambda aligns itself with the scalar field $\lambda=\sqrt{\rho}\psi$ (alignment of density \textit{and} phase). In what physical scenario does the vacuum essentially align itself with the phase of $\psi$? Are there any other expression (i.e. special scenarios) one can articulate from such coupled equations of motion? 

The non-local theory presented in the past two sections has been shown to be a subset of the coupled 2nd-order equations (Eq.~\ref{WFeqnVacuumF1}-~\ref{lambdaPsi_v3}). The presented coupled equations of motion, with the inclusion of two fields, appears to contain a rich collection of physical scenarios worth further exploring. We therefore propose the coupled equations of motion (Eq.~\ref{LambdaNiceEqExp} and Eq.~\ref{WFeqnVacuumFExp}), derived from a geometric coupling of General Relativity and quantum mechanics~\cite{our_cosmo_paper}, to be insightful in characterizing field equations for particles embedded in a quantum vacuum
\begin{eqnarray}
\Omega^2\mathcal{D}_{\mu}\mathcal{D}^{\mu}\psi 
- \Bigl[\frac{1}{2m\rho}\Bigl(\Box
- \frac{2m^2}{\hbar^2}(1-Q)\Bigr)\lambda\Bigr]\psi =0 \label{WFeqnVacuumFExp}. 
\end{eqnarray}

\begin{eqnarray}
\frac{m^2}{\hbar^2}(1-Q)\lambda=\nabla_{\mu}\Bigl(\lambda\frac{\nabla^{\mu}\sqrt{\rho}}{\sqrt{\rho}} \Bigr) \label{LambdaNiceEqExp}
\end{eqnarray}
The above coupled equations correspond to the yet unexplored exponential constraint $\Omega^2=e^{Q}$. Currently, these equations apply to one particle. A Many-Body extension requires a deeper understanding of how the kinetic energy of one particle implicates another (if at all), and hence, further investigation is required. 
We believe that these equations can serve as a path for attaining the yet unattained higher-order contributions in quantum mechanics. 
These higher-order contributions are deeply tied to the vacuum and we therefore emphasize the importance of the vacuum field $\lambda$ and its associated equation (Eq.~\ref{LambdaNiceEqExp}).

\section{Conclusion}
In this manuscript, matter is manifestly embedded within a conformally transformed spacetime to account for its interaction with its own internal kinetic energy. 
To alleviate the need for a continuity equation, a second-order wavefunction equation, coupled to the acclaimed vacuum field $\lambda$, was derived from its equivalent Bohmian framework.
A special case, whereby the vacuum field aligns itself with the KG field $\lambda=\sqrt{\rho}\psi$ has been explored and various implications were given thereof. 
Modified canonical commutation and energy-momentum relations were defined for a null vacuum-phase scenario and a representation theorem, between momentum and differential form operators, was introduced.
It was further shown that such commutation relations eliminate infrared and ultraviolet divergences in the associated Klein-Gordon propagator. Further work is needed to explore the second quantized version of such an extended quantum theory.  
Characterizing a Lagrangian for the proposed nonlinear, coupled field equations is nontrivial, even when the fields $\lambda$ and $\psi$ conform. 
Regardless, it is our hope that, in incorporating the interaction of the particle to its own internal kinetic energy, we can move one step closer to unifying the two theoretical frameworks in a more coherent manner. 

\bibliographystyle{JHEP}
\bibliography{quantum_gravity_bibtex_link}

\providecommand{\href}[2]{#2}\begingroup\raggedright\begin{thebibliography}{10}

\bibitem{QMGeometry}
F.~{Shojai} and A.~{Shojai}, \emph{{Understanding Quantum Theory in Terms of
  Geometry}}, {\emph{ArXiv e-prints} (2004) }
  [\href{https://arxiv.org/abs/gr-qc/0404102}{{\ttfamily gr-qc/0404102}}].

\bibitem{our_cosmo_paper}
D.~{Gabay} and S.~K. {Joseph}, \emph{{On the Cosmological Constant in a
  Conformally Transformed Einstein Equation}}, {\emph{ArXiv e-prints} (2018) }
  [\href{https://arxiv.org/abs/1801.00161}{{\ttfamily 1801.00161}}].

\bibitem{Sidharth_QM}
B.~Sidharth, \emph{Geometry and quantum mechanics}, {\emph{Ann. Fond. Louis de
  Broglie} {\bfseries 29} (2004) }
  [\href{https://arxiv.org/abs/physics/0211012}{{\ttfamily physics/0211012}}].

\bibitem{QRGM_Susskind}
L.~{Susskind}, \emph{{Dear Qubitzers, GR=QM}}, {\emph{ArXiv e-prints} (2017) }
  [\href{https://arxiv.org/abs/1708.03040}{{\ttfamily 1708.03040}}].

\bibitem{BohmI}
D.~Bohm, \emph{A suggested interpretation of the quantum theory in terms of
  "hidden" variables. i},
  \href{https://doi.org/10.1103/PhysRev.85.166}{\emph{Phys. Rev.} {\bfseries
  85} (1952) 166}.

\bibitem{BohmII}
D.~Bohm, \emph{A suggested interpretation of the quantum theory in terms of
  "hidden" variables. ii},
  \href{https://doi.org/10.1103/PhysRev.85.180}{\emph{Phys. Rev.} {\bfseries
  85} (1952) 180}.

\bibitem{PHollandBook}
P.~R. Holland, \emph{The Quantum Theory of Motion}. Cambridge University Press,
  Cambridge, United Kingdom, first edition~ed., 1995.

\bibitem{SheldonReview}
S.~{Goldstein}, \emph{{Bohmian Mechanics and the Quantum Revolution}},
  {\emph{eprint arXiv:quant-ph/9512027} (1995) }
  [\href{https://arxiv.org/abs/quant-ph/9512027}{{\ttfamily
  quant-ph/9512027}}].

\bibitem{Shojai_Article}
F.~Shojai and M.~Golshani, \emph{On the geometrization of bohmian mechanics: A
  new approach to quantum gravity},
  \href{https://doi.org/10.1142/S0217751X98000305}{\emph{Int. J. Mod. Phys. A}
  {\bfseries 13} (1998) 677}.

\bibitem{ShojaiBohmianQM}
A.~Shojai and F.~Shojai, \emph{About some problems raised by the relativistic
  form of de-broglie-bohm theory of pilot wave}, {\emph{Phys. Scripta}
  {\bfseries 64} (2001) 413}
  [\href{https://arxiv.org/abs/quant-ph/0109025}{{\ttfamily
  quant-ph/0109025}}].

\bibitem{ShojaiCnstrAlgebra}
A.~Shojai and F.~Shojai, \emph{Constraint algebra and equations of motion in
  the bohmian interpretation of quantum gravity}, {\emph{Class. Quantum Grav.}
  {\bfseries 21} (2004) 1}
  [\href{https://arxiv.org/abs/gr-qc/0311076}{{\ttfamily gr-qc/0311076}}].

\bibitem{Shojai_ScalarTensor}
F.~Shojai and A.~Shojai, \emph{Nonminimal scalar-tensor theories and quantum
  gravity}, \href{https://doi.org/10.1142/S0217751X0000080X}{\emph{Int. J. Mod.
  Phys. A} {\bfseries 15} (2000) 1859}
  [\href{https://arxiv.org/abs/gr-qc/0010012}{{\ttfamily gr-qc/0010012}}].

\bibitem{Bender1}
C.~M. Bender and P.~D. Mannheim, \emph{$\mathcal{P}\mathcal{T}$ symmetry and
  necessary and sufficient conditions for the reality of energy eigenvalues},
  \href{https://doi.org/10.1016/j.physleta.2010.02.032}{\emph{Phys. Lett. A}
  {\bfseries 374} (2010) 1616 }.

\bibitem{Bender2}
C.~M. Bender and P.~D. Mannheim, \emph{Exactly solvable
  $\mathcal{P}\mathcal{T}$-symmetric hamiltonian having no hermitian
  counterpart}, \href{https://doi.org/10.1103/PhysRevD.78.025022}{\emph{Phys.
  Rev. D} {\bfseries 78} (2008) 025022}.

\bibitem{Bender3}
C.~M. Bender and P.~D. Mannheim, \emph{No-ghost theorem for the fourth-order
  derivative pais-uhlenbeck oscillator model},
  \href{https://doi.org/10.1103/PhysRevLett.100.110402}{\emph{Phys. Rev. Lett.}
  {\bfseries 100} (2008) 110402}.

\bibitem{Bender4}
C.~M. Bender and P.~D. Mannheim, \emph{$\mathcal{P}\mathcal{T}$ symmetry in
  relativistic quantum mechanics},
  \href{https://doi.org/10.1103/PhysRevD.84.105038}{\emph{Phys. Rev. D}
  {\bfseries 84} (2011) 105038}.

\bibitem{Manheim_PTSym}
P.~D. Mannheim, \emph{$\mathcal{P}\mathcal{T}$ symmetry as a necessary and
  sufficient condition for unitary time evolution},
  \href{https://doi.org/10.1098/rsta.2012.0060}{\emph{Philos. Trans. Royal Soc.
  A} {\bfseries 371} (2013) }.

\bibitem{WeinbegCosmo89}
S.~Weinberg, \emph{The cosmological constant problem},
  \href{https://doi.org/10.1103/RevModPhys.61.1}{\emph{Rev. Mod. Phys.}
  {\bfseries 61} (1989) 1}.

\bibitem{Padic_MathPhys}
B.~Dragovich, A.~Y. Khrennikov, S.~V. Kozyrev, I.~V. Volovich and E.~I.
  Zelenov, \emph{p-adic mathematical physics: the first 30 years},
  {\emph{p-Adic Numbers, Ultrametric Analysis, \& Applications} {\bfseries 9}
  (2017) 87}.

\bibitem{Calcagni2010}
G.~Calcagni and G.~Nardelli, \emph{String theory as a diffusing system},
  \href{https://doi.org/10.1007/JHEP02(2010)093}{\emph{J. High Energy Phys.}
  (2010) 93} [\href{https://arxiv.org/abs/0910.2160}{{\ttfamily 0910.2160}}].

\bibitem{RollingTachyon}
N.~Moeller and B.~Zwiebach, \emph{Dynamics with infinitely many time
  derivatives and rolling tachyons},
  \href{https://doi.org/10.1088/1126-6708/2002/10/034}{\emph{J. High Energy
  Phys.} {\bfseries 2002} (2002) 034}
  [\href{https://arxiv.org/abs/hep-th/0207107}{{\ttfamily hep-th/0207107}}].

\bibitem{QMSourceBook}
B.~L. V.~D. Waerden, \emph{Sources of Quantum Mechanics}. Dover Publications,
  New York, 1967.

\bibitem{SidharthLambRevw}
B.~Sidharth, A.~Das and A.~D. Roy, \emph{Report: Anomalous gyromagnetic
  ratio,revisiting the lamb shift, lorentz invariance violation ...},
  {\emph{New Adv. Phys.} {\bfseries 9} (2015) 67}.

\bibitem{Freund87padic}
P.~G. Freund and M.~Olson, \emph{Non-archimedean strings},
  \href{https://doi.org/10.1016/0370-2693(87)91356-6}{\emph{Physics Letters B}
  {\bfseries 199} (1987) 186 }.

\bibitem{Zwiebach_String2002}
N.~{Moeller} and B.~{Zwiebach}, \emph{{Dynamics with Infinitely Many Time
  Derivatives and Rolling Tachyons}}, {\emph{J. High Energy Phys.} {\bfseries
  10} (2002) 034} [\href{https://arxiv.org/abs/hep-th/0207107}{{\ttfamily
  hep-th/0207107}}].

\bibitem{Gianluca2008}
G.~Calcagni and G.~Nardelli, \emph{Tachyon solutions in boundary and open
  string field theory},
  \href{https://doi.org/10.1103/PhysRevD.78.126010}{\emph{Phys. Rev. D}
  {\bfseries 78} (2008) 126010}
  [\href{https://arxiv.org/abs/0708.0366}{{\ttfamily 0708.0366}}].

\bibitem{OstroInstab2015}
H.~Motohashi and T.~Suyama, \emph{Third order equations of motion and the
  ostrogradsky instability},
  \href{https://doi.org/10.1103/PhysRevD.91.085009}{\emph{Phys. Rev. D}
  {\bfseries 91} (2015) 085009}
  [\href{https://arxiv.org/abs/1411.3721}{{\ttfamily 1411.3721}}].

\bibitem{Woodard2007Ostro}
R.~{Woodard}, \emph{{Avoiding Dark Energy with $1/R$ Modifications of
  Gravity}},  in \emph{The Invisible Universe: Dark Matter and Dark Energy}
  (L.~{Papantonopoulos}, ed.), vol.~720 of \emph{Lecture Notes in Physics,
  Berlin Springer Verlag}, p.~403, 2007,
  \href{https://arxiv.org/abs/astro-ph/0601672}{{\ttfamily astro-ph/0601672}}.

\bibitem{Woodard2015Ostro}
R.~P. {Woodard}, \emph{{The Theorem of Ostrogradsky}}, {\emph{ArXiv e-prints}
  (2015) } [\href{https://arxiv.org/abs/1506.02210}{{\ttfamily 1506.02210}}].

\bibitem{BenderManheimGhost08}
C.~M. Bender and P.~D. Mannheim, \emph{Giving up the ghost},
  \href{https://doi.org/10.1088/1751-8113/41/30/304018}{\emph{J. Phys. A}
  {\bfseries 41} (2008) 304018}
  [\href{https://arxiv.org/abs/0807.2607}{{\ttfamily 0807.2607}}].

\bibitem{Chen2013ostro}
T.-j. {Chen}, M.~{Fasiello}, E.~A. {Lim} and A.~J. {Tolley}, \emph{{Higher
  derivative theories with constraints: exorcising Ostrogradski's ghost}},
  {\emph{J. Cosmol. Astropart. Phys.} {\bfseries 2} (2013) 042}
  [\href{https://arxiv.org/abs/1209.0583}{{\ttfamily 1209.0583}}].

\bibitem{BarnabyKamranOstro}
N.~{Barnaby} and N.~{Kamran}, \emph{{Dynamics with infinitely many derivatives:
  the initial value problem}}, {\emph{J. High Energy Phys.} {\bfseries 2}
  (2008) 008} [\href{https://arxiv.org/abs/0709.3968}{{\ttfamily 0709.3968}}].

\end{thebibliography}\endgroup

\end{document}